\documentclass[superscriptaddress,prd,showpacs,twocolumn,nofootinbib]{revtex4-2}
\usepackage{graphicx}
\usepackage{xcolor}
\usepackage{epsf}
\usepackage{bm}
\usepackage{amsmath}
\usepackage{amsfonts}
\usepackage{amssymb}
\usepackage{epstopdf}
\usepackage{makecell}
\usepackage{booktabs}
\usepackage[linktocpage]{hyperref}
\usepackage{hyperref}
\usepackage{soul}
\usepackage{ulem}
\hypersetup{colorlinks=true, citecolor=red, linkcolor=blue,
	urlcolor = magenta, filecolor=magenta}
\def\cL{\mathcal{L}}
\def\be{\begin{equation}}
\def\ee{\end{equation}}
\def\bea{\begin{eqnarray}}
\def\eea{\end{eqnarray}}

\begin{document}

\title{The Weyl Geometric Gravity black hole in light of the Solar System tests}
	\author{Mohsen Khodadi}
	\email{khodadi@kntu.ac.ir}
	\affiliation{School of Physics, Damghan University, Damghan 3671641167, Iran}
	\affiliation{Center for Theoretical Physics, Khazar University, 41 Mehseti Street, AZ1096 Baku, Azerbaijan}
\author{Tiberiu Harko}
\email{tiberiu.harko@aira.astro.ro}
\affiliation{Faculty of Physics, Babe\c s-Bolyai University,
	1 Kog\u alniceanu Street, Cluj-Napoca 400084, Romania}	
\affiliation{Astronomical Observatory, 19 Cire\c silor Street, Cluj-Napoca 400487, Romania}

\begin{abstract}
	 The Weyl geometric gravity theory, in which the gravitational action is constructed from the square of the Weyl curvature scalar, and \st{of} the strength of the Weyl vector, has been intensively investigated recently. The theory admits a scalar-vector-tensor representation, obtained by introducing an auxiliary scalar field, and can therefore be reformulated as a scalar-vector-tensor theory in a Riemann space, in the presence of a nonminimal coupling between the Ricci scalar and the scalar field. By assuming that the Weyl vector has only a radial component, an exact spherically symmetric vacuum solution of the field equations can be obtained, which depends on three integration constants. As compared to the Schwarzschild solution, the Weyl geometric gravity solution contains two new terms, linear and quadratic in the radial coordinate, respectively.  In the present work we consider the possibility of testing and obtaining observational restrictions on the Weyl geometric gravity black hole at the scale of the Solar System, by considering six classical tests of general relativity (gravitational redshift, the E\"{o}tv\"{o}s parameter and the universality of free fall, the Nortvedt effect, the planetary perihelion precession, the deflection of light by a compact object, and the radar echo delay effect, respectively) for the exact spherically symmetric black hole solution of the Weyl geometric gravity. All these gravitational effects can be fully explained and are consistent with the vacuum solution of the Weyl geometric gravity. Moreover, the study of the classical general relativistic tests also allows to constrain the free parameter of the solution.
	\end{abstract}
	\maketitle

\color{blue}{\tableofcontents}

\color{black}

\section{Introduction}\label{sec:intro}

Weyl's generalization of Riemann geometry has attracted a lot interest in both fields of mathematics and physics \cite{Weyl1,Weyl2, Weyl3,Weyl4}.  One of the fundamental properties of the Riemannian geometry, and consequently of Einstein's general relativity, is that the transport of vector lengths is integrable. In Weyl’s geometry, when transported along closed curves, the lengths of vectors change. Therefore, in Weyl geometry, parallel transport takes into account the local properties of spacetime, which leads to the fundamental property of this geometry, namely, that under parallel transport the lengths of vectors is non-integrable. The Riemannian geometry was generalized by Weyl through the introduction of a new geometrical degree of freedom, the Weyl vector $\omega _\mu$. 

In Weyl geometry the covariant derivative $\nabla _\mu$ of the metric tensor does not vanish identically, and this property leads to the fundamental geometric concept of non-metricity $Q_{\lambda \mu \nu}$, defined through the Weyl compatibility condition $\nabla _\lambda g_{\mu \nu}= Q_{\lambda \mu \nu}$. Another fundamental concept introduced by Weyl, which has important physical implications, is the notion of conformal invariance. 

Weyl also proposed that all laws of physics must be invariant with respect to conformal transformations, that is, they are unchanged under local conformal transformations of the form  $d\tilde{s}^2 = \Psi ^n(x)ds^2 = \Psi ^n(x)g_{\mu \nu}dx^\mu dx^\nu =  \tilde{g}_{\mu \nu}dx^\mu dx^\nu$. The conformal transformations relate the changes in the units of length and time at each point of the spacetime. $\Psi (x)$ is called the conformal factor, and $n$ is the Weyl charge.

Despite the severe criticism by Einstein, Weyl's geometry was intensively applied, and investigated from the point of view of physical applications. Dirac \cite{Di1,Di2} generalized Weyl's theory by introducing a real scalar field $\phi$ of weight $w(\phi)=-1$. For the description of the gravitational interaction Dirac introduced the Lagrangian $L=-\phi^2R +kD_\mu \phi D^\mu \phi+c\phi^4 + (1/4)F_{\mu \nu}F^{\mu \nu}$, where $R$ is the Ricci scalar, and $c$ and $k = 6$ are constants. By $F_{\mu \nu}$ we have denoted the electromagnetic type tensor defined with the help of the Weyl vector. By construction the Dirac Lagrangian is conformally invariant. 

The principle of conformal invariance was fully implemented in a class of theories known as conformal gravity \cite{Ma1,Ma2}, based on the action $I_W=-\alpha _g\int{C_{\lambda \mu \nu \sigma}C^{\lambda \mu \nu \sigma}\sqrt{-g}d^4x}$, where $\alpha_g$ is a coupling constant,  and $C_{\lambda \mu \nu \sigma}$ is the Weyl conformal tensor. Conformal gravity can give a gravitational explanation of the dark matter problem without the need of introducing an unknown form of matter. Weyl geometry also represents the mathematical foundation of the $f(Q)$ theory of gravity, and of its generalizations \cite{Q1,Q2,Q3,Q4,Q5,Q6,Q7}, based on the fundamental action $S=\int{f(Q)\sqrt{-g}d^4x}$, where $Q$ is the nonmetricity scalar. For detailed investigations of various aspects and physical implications of Weyl geometry and Weyl quadratic gravity see \cite{D1,D2,D3,D4,D5}.

A novel view of Weyl quadratic gravity, based on the action $S_W=\int{\tilde{R}^2\sqrt{-g}d^4x}$, where $\tilde{R}$ is the Weyl scalar,  was proposed in \cite{Gh1,Gh2,Gh3,Gh4,Gh5,Gh6,Gh7,Gh8,Gh9,Gh10,Gh11,Gh12, Gh13,Gh14,Gh15,Gh16}, and is based on the linearization of $S_W$ by introducing an auxiliary scalar field $\phi$.  This minimal approach does not introduce any additional degrees of freedom in the theory, and the Weyl gauge symmetry $D(1)$ is broken spontaneously through a geometric Stueckelberg mechanism. This leads to an Einstein-Proca type action, in which the Weyl gauge field $\omega _\mu$ is naturally included. We will call in the following the scalar-vector-tensor gravitational theory based on the linearization of the Weyl quadratic Lagrangian as the Weyl Geometric Gravity (WGG) theory.

Various cosmological and astrophysical implications of the WGG theory have been investigated recently. The coupling of matter to geometry in conformal quadratic Weyl gravity, by assuming a coupling term of the form $L_m\tilde{R}^2$, where $L_m$ is the matter Lagrangian, and $\tilde{R}$ is the Weyl scalar, was investigated in \cite{Ha1}. The cosmological applications of the theory were also considered, and it was shown that the model can give a good description of the observational data for the Hubble function up to a redshift of the order of $z\approx 3$. Exact and numerical black hole solutions in the WGG theory were considered in \cite{Yang:2022icz}. The possibility of explaining dark matter as a Weyl geometric effect was considered in \cite{Burikham:2023bil}.  The structure and physical properties of specific classes of neutron, quark, and Bose-Einstein condensate stars in the conformally invariant WGG theory were analyzed in \cite{Ha2}.   As a general result it was found that several equations of state of high density matter, Weyl geometric gravity stars are more massive than their general relativistic counterparts. 

The effects of Weyl geometry on the propagation of electromagnetic wave packets, and on the gravitational spin Hall effect of light, were studied in \cite{Ha3}. The thermodynamic properties of an exact black hole solution obtained in WGG theory were investigated in \cite{Ha4}. The Weyl geometric black holes have thermodynamic properties that significantly differentiate them from similar solutions of general relativity and of other modified gravity theories. These differences could lead to the possibility of a better understanding of the properties of the black holes in alternative theories of gravity, and of the relevance of the thermodynamic aspects in black hole physics. The cosmological implications of the WGG theory were further investigated in \cite{Ha5}. Two cosmological models, corresponding to the vacuum state, and to the presence of matter described by a linear barotropic equation of state, were investigated.  A mimetic extension of the WGG theory was introduced in \cite{Ha6}. 

A detailed investigation of the properties of the galactic rotation curves in the WGG theory was presented in \cite{Ha7}. The theoretical predictions were tested by using 175 galaxies from the Spitzer Photometry \& Accurate Rotation Curves (SPARC) database.  The exact solution of the Weyl geometric gravity can successfully describe a the large variety of the rotation curves of the SPARC sample, and thus give a satisfactory description of the particle motion in the galactic halos, without the need of introducing an extra dark matter component. 

The astrophysical properties of the exact black hole solution obtained in Weyl geometric gravity theory were considered in \cite{Ha8}, by performing a detailed analysis of its lensing properties in the strong field regimes. The shadow of the black hole was analyzed, leading to a first set of constraints on the solution parameters by using the observational data from the shadows of the M87* and Sgr A* supermassive black holes. The strong lensing in this geometry was also investigated,  including the study of the angular position of the images, the angular separation, the relative magnification, the radii of the Einstein ring, and the relativistic time delay. 

One of its most important predictions of general relativity is the existence of black holes. Two basic predictions related to the existence of these mysterious cosmic objects have been recently confirmed.   The first detection by the LIGO and VIRGO of the gravitational waves \cite{grav1,grav2} provided direct evidence for the existence of black holes. Moreover, the existence of black holes was visually confirmed by the Event Horizon Telescope (EHT) collaboration, which  presented the first image of the plasma orbiting the black hole at the center of galaxy M87 \cite{grav3}.  EHT also obtained an image of Sagittarius A* (Sgr A*), the supermassive black hole at center of the Milky Way galaxy \cite{grav4}. These discoveries confirmed the predictions of general relativity, and they have opened new avenues for the  understanding of the compact astrophysical objects \cite{N0a,N0b,N0c,N1,N2,N3,N4}. 

Hence, the properties of black holes, representing exact solutions of the vacuum gravitational field equations, are powerful indicators of the physical viability of the given theory, as well as an important testing ground for the mathematical structure of the theory. For the Weyl geometric gravity an exact solution of the vacuum gravitational field equations in static spherical symmetry was obtained in \cite{Yang:2022icz} and \cite{Burikham:2023bil}. The solution represents a generalization of the Schwarzschild - de Sitter solution of standard general relativity, through the presence of two new terms in the metric, having a linear and a quadratic dependence on the radial coordinate, respectively. The solution is obtained by assuming that the Weyl vector has only a radial component. The solution is three-parametric, depending on three arbitrary integration constants. From physical considerations two of these constants can be fixed as the gravitational radius (mass) of the central compact object, and the cosmological constant, respectively. Hence, the solution depends essentially on only one free parameter, which enters in the metric as a constant term, and as determining the term linearly increasing with $r$. We will call in the following this solution as the WGG black hole.  The physical properties and the astrophysical implications of the WGG black hole were investigated from various points of view in \cite{Ha3,Ha4} and \cite{Ha7,Ha8}, respectively. 

It is the goal of the present paper to perform a detailed investigation of the WGG black hole solution, by considering the Solar System tests of the solution. These are the classical tests of general relativity, based on very precise observations, and which provide very sharp limits on the free parameters of the vacuum static solutions of the gravitational theories.            

In our analysis of the metric of the WGG black hole we will consider six classical tests of GR, namely, the gravitational redshift, the Weak Equivalence Principle (WEP), the Strong Equivalence Principle (SEP), the perihelion precession, the light deflection, and the radar echo delay, respectively. These gravitational effects allow to obtain an upper bound for the dimensionless free parameter of the WGG solution $\eta$, which is obtained as $|\eta|<10^{-10}$. It is interesting to point out that from the analysis of the galactic rotation curves in this metric, performed in \cite{Ha7}, a value of $\eta$ of the order of $\eta \approx 10^{-15}$ was determined. Hence the WGG black hole solution passes the astrophysical tests at both galactic and Solar System level.

The exploration of gravitational theories based on non-metricity, such as 
$f(Q)$ gravity, continues to be a vibrant area of research, with recent studies providing significant cosmological constraints. For instance, phenomenological aspects of logarithmic $f(Q)$ models have been investigated using latest-generation data \cite{Karmakar:2025iba}, and the degeneracy between dynamical dark energy and modified gravity has been analyzed through signatures in gravitational wave propagation \cite{Karmakar:2025yng}. These works highlight the importance of combining geometric formulations of gravity with high-precision observational data across different scales, a philosophy that also motivates the present Solar System test of Weyl Geometric Gravity.

The present paper is organized as follows. The WGG black hole solution is introduced in Section~\ref{sect1}. A first class of observational tests - gravitational redshift, the universality of the free fall, and the Nordvedt effect are considered in Section~\ref{sect2}. A second class of tests - perihelion precession, light bending, and the radar echo delay are investigated in Section~\ref{sect3}, leading to the determination of sharp upper limits for the free parameter $\eta$ of the solution. Finally, we discuss and conclude our results in Section~\ref{sect4}.        
	
\section{Exact black hole solution in Weyl geometric gravity}\label{sect1}

In this Section, by following the results of \cite{Yang:2022icz,Burikham:2023bil}, we briefly introduce the exact black hole solution of the Weyl geometric gravity theory. 
As for the space-time geometry, we will consider it to be static and spherically symmetric, with the metric represented by  
\begin{equation}
ds^2= e^{\nu(r)}dt^2-e^{\lambda(r)}dr^2-r^2(d\theta^2+\sin^2\theta d\varphi^2),
\end{equation} 
where we assume $\lambda(r)=-\nu(r)$, and the metric is written down in the coordinates $\left(x^0=ct,x^1=r,x^2=\theta, x^3=\varphi\right)$.  The Weyl geometric gravity theory is based on the action \cite{Yang:2022icz,Burikham:2023bil}
\bea\label{act}
\mathcal{S} &=& \int d^4x \sqrt{-g} \Bigg[ -\frac{1}{12} \frac{\Phi}{\xi^2} \left( R - 3\alpha \nabla_\mu \omega^\mu - \frac{3}{2} \alpha^2 \omega_\mu \omega^\mu \right)\nonumber\\
&& - \frac{1}{4!} \frac{\Phi^2}{\xi^2} - \frac{1}{4} F_{\mu\nu} F^{\mu\nu} \Bigg].
\eea

This action extends standard general relativity through the inclusion of two new fields, a scalar  $\Phi$, and of the Weyl vector field $\omega_{\mu}$, in addition to the metric field $g_{\mu\nu}$. Here $\xi$ is a dimensionless perturbative coupling constant,  satisfying the restriction  $\xi<1$, and $\alpha$ is the dimensionless Weyl gauge coupling constant,  which is introduced together with  the nonmetricity condition $\tilde{\bigtriangledown}_\lambda g_{\mu\nu}=-\alpha \omega_{\lambda}g_{\mu\nu} $, which represents a key feature of Weyl gravity. 

In action (\ref{act}) $F_{\mu\nu}$ is the Weyl field strength, defined in the form $\tilde{F}_{\alpha\beta} = \tilde{\nabla}_{[\alpha} \omega_{\beta]} = \nabla_{[\alpha} \omega_{\beta]} = \partial_{[\alpha} \omega_{\beta]} = \partial_\alpha \omega_\beta - \partial_\beta \omega_\alpha$.

The action (\ref{act}) is obtained from the linearization of the  most general gravitational Lagrangian density defined in a Weyl geometry, and which is invariant under a gauged Weyl symmetry, given by \cite{Gh6,Gh7,Gh8}
\bea\label{L1}
\cL_1=\sqrt{g}\,\,\Big\{\frac{1}{4!\,\xi^2} \tilde R^2-\frac{1}{\eta^2}\, \tilde C_{\mu\nu\rho\sigma}^2
-\frac{1}{4}\,F_{\mu\nu}^2\Big\}, 0<\xi,\,\eta <1, \nonumber\\
\eea
where $\tilde R$ is the Weyl scalar curvature, and $\tilde C_{\mu\nu\rho\sigma}^2$ is the Weyl tensor, whose effects we will neglect in our present approach. This is equivalent in taking the limit $\eta \rightarrow \infty$ in (\ref{L1}). To obtain the scalar-vector-tensor representation of the action (\ref{L1}) we perform the substitution \cite{Gh6,Gh7,Gh8}
\be
\tilde R^2\rightarrow - 2 \phi^2 \tilde R-\phi^4,
\ee
where $R$ is the Riemannian scalar curvature, where $\phi$ is an auxiliary scalar field.  In this way we obtain the action (\ref{act}) of the Weyl geometric gravity in its scalar-vector-tensor representation. 

By varying the action (\ref{act}) with respect to the metric tensor we find the Einstein equation  of the Weyl geometric gravity theory as
\bea\label{field}
&&\frac{\Phi}{\xi^2} \left( R_{\mu\nu} - \frac{1}{2} R g_{\mu\nu} \right) - \frac{3\alpha}{2\xi^2} \left(g_{\mu\nu} \omega^\rho \nabla_\rho   - \omega_\nu \nabla_\mu  - \omega_\mu \nabla_\nu  \right)\Phi \nonumber\\
&& + \frac{3\alpha^2}{4\xi^2} \Phi \left( \omega_\rho \omega^\rho g_{\mu\nu} - 2 \omega_\mu \omega_\nu \right) + 
6 F_{\rho\mu} F_{\sigma\nu} g^{\rho\sigma} - \frac{3}{2} F_{\rho\sigma}^2 g_{\mu\nu} \nonumber\\
&&- \frac{1}{4\xi^2} \Phi^4 g_{\mu\nu} + \frac{1}{\xi^2} \left( g_{\mu\nu} \Box - \nabla_\mu \nabla_\nu \right) \Phi = 0.
\eea
By taking the trace of the field equation we find
\begin{equation}
\Phi R + 3\alpha \omega^\rho \nabla_\rho \Phi + \Phi^2 - \frac{3}{2} \alpha^2 \Phi \omega_\rho \omega^\rho - 3 \Box \Phi = 0
\end{equation}

The variation of the action (\ref{act}) with respect to the scalar field $\Phi$ and of the Weyl vector field $\omega_{\mu}$ gives   
\begin{equation}
R - 3\alpha \nabla_\rho \omega^\rho - \frac{3}{2} \alpha^2 \omega_\rho \omega^\rho + \Phi = 0,
\end{equation}
and the  generalized Klein–Gordon (GKG) 
\begin{equation}\label{p}
\Box \Phi - \alpha \nabla_\rho (\Phi \omega^\rho) = 0,
\end{equation}
and Maxwell type equations
\begin{equation}\label{f}
4\xi^2 \nabla_\nu F^{\mu\nu} - \alpha^2 \Phi \omega^\mu + \alpha \nabla^\mu \Phi = 0,
\end{equation}
respectively.

The exact vacuum solution of the Weyl geometric gravity field equations (\ref{field}) can be obtained by assuming the following ansatz for the Weyl vector 
\begin{equation}
\omega_\mu = (0, \omega_1(r), 0, 0),
\end{equation} 
which gives the following form of the metric fully satisfying the field equations Eqs.~(\ref{field})-(\ref{f}) 
\begin{equation}\label{metric}
f(r)=e^{\nu (r)} = e^{-\lambda (r)} = 1 - \eta + \frac{\eta(2 - \eta)}{3} \frac{r}{r_g} - \frac{r_g}{r} -\frac{\Lambda}{3} r^2,
\end{equation}
where $\eta $, $r_g$ and $\Lambda$ are arbitrary integration constants \cite{Yang:2022icz,Burikham:2023bil}.  By comparing the above solution with the Schwarzschild metric, one finds the interpretation of the constant $r_g$ as the gravitational radius $r_g=2M$.

Eq.~(\ref{metric}) represents  the exact static spherically symmetric black hole solution in Weyl geometric gravity with a radial Weyl vector. It is clear that by setting $\eta=0$, we recover the Schwarzschild-de Sitter metric. 

\section{Classical tests of Weyl geometric gravity-gravitational redshift, WEP, and SEP}\label{sect2}

We begin our analyses of the six classical tests of GR for the Weyl geometric gravity black hole by considering the  Gravitational Redshift, the Weak Equivalence Principle (WEP), and the Strong Equivalence Principle (SEP), respectively. These effects will allow us to obtain a first set of constraints on the free parameter $\eta$ of the solution. For the analysis conducted in the rest of the paper we adopt the metric convention $(-,+,+,+)$.

\subsection{Gravitational redshift}

The standard GR expression for the redshift $z$ between two points $r_1$ and $r_2$ in the Schwarzschild metric, corresponding to the limit  $\eta=0$, $\Lambda=0$ of the Weyl geometric gravity metric, is given by  
\begin{equation}
(1+z)_{\text{GR}} = \sqrt{\frac{1 - \frac{2GM}{c^2 r_1}}{1 - \frac{2GM}{c^2 r_2}}}. 
\end{equation}

The expression of the redshift for the WGG metric is
\begin{equation}\label{wgg}
(1+z)_{\text{WGG}} = \sqrt{ \frac{ 1 - \eta + \dfrac{\eta(2 - \eta)}{6M}r_1 - \dfrac{2GM}{c^2 r_1} } { 1 - \eta + \dfrac{\eta(2 - \eta)}{6M}r_2 - \dfrac{2GM}{c^2 r_2} } }
\end{equation}

There are two new terms appearing in the redshift for WGG black holes. The term ($1-\eta$) represents a constant shift with respect to the Newtonian potential. It affects the redshift even at finite distances and would be present even if the two measurement points were at the same gravitational potential (e.g., same $r$ value) in the Schwarzschild sense.
The term $[\eta(2-\eta)/6M]r$ grows linearly with $r$. Its effect becomes more significant for experiments conducted over larger spatial scales.
Any precise measurement of $z$ that matches the GR prediction would force these extra terms to be negligible, thereby constraining \(|\eta|\) to be very small.

The most stringent constraints come from experiments where the predictions of GR are confirmed with high precision. Hence the logic of our investigation is to assume that the measured redshift is consistent with GR, and then to see what values of $\eta$ would cause a deviation larger than the experimental error bar. An experiment measures a value $z_{\text{exp}}$ with an uncertainty $\delta z$, meaning we assume the true value is \(z_{\rm exp}=z_{\rm GR}(\eta=0)\). The difference between the prediction with $\eta\neq 0 $ and the GR prediction must be smaller than the experimental uncertainty
\begin{equation}\label{in}
 \left| z_{\text{WGG}} - z_{GR} \right| < \delta z. 
 \end{equation}

Solving this inequality will yield an allowed range for $\eta$, e.g., $-\eta_{\text{max}} < \eta < \eta_{\text{max}}$.

The Vessot-Levine rocket experiment, also called the Gravity Probe A (GP-A) experiment \cite{Vessot:1979,Vessot:1980zz} is the classic and most precise test of gravitational redshift. A hydrogen maser clock on a rocket was launched to an altitude of about $10^4$ km ($r_2 \approx R_\oplus + 10^4$ km). Its frequency was compared to a similar clock on Earth ($r_1 = R_\oplus$). 
The experiment confirmed the GR prediction to a precision of $\delta z / z \approx 1.4 \times 10^{-4}$ (with a 0.014\% accuracy). 

Let's analyze such an experiment within Earth's gravity, i.e., $r_1 = R_\oplus$ represents the radial coordinate of the clock on the Earth's surface,  and $r_2 = R_\oplus + h$ is the radial coordinate of the clock at an altitude above the Earth's surface. For such an experiment within Earth's gravity, the $\Lambda$ term is utterly negligible. The linear term $[(\eta(2-\eta)/6M]r$ is also very small for Earth's mass and its radius. The dominant new effect is the constant offset $(1-\eta)$.

Let's simplify the redshift formula for this case. We expand the square roots in the weak field limit ($2GM/c^2r << 1$). The explicit calculation of the terms involving $\eta$ in the weak-field expansion of the gravitational redshift formula can be done as follows. We start with the full expression (\ref{wgg}) of the gravitational redshift. 
Let's denote the function under the square root at a general point $r$ as
\be
f(r) = 1 - \eta - \frac{2GM}{c^2 r} + \frac{\eta(2 - \eta)}{6M}r.
\ee
We work in the weak-field regime where all terms after the $1$ are small. Let us introduce the quantity $\alpha (r)$, defined as
\be
\alpha(r) = -\eta - \frac{2GM}{c^2 r} + \frac{\eta(2 - \eta)}{6M}r.
\ee
so that $f(r) = 1 + \alpha(r)$, with $|\alpha(r)| \ll 1$.

Now, we can expand the square root for both numerator and denominator as
\be
\sqrt{f(r)} = \sqrt{1 + \alpha(r)} \approx 1 + \frac{1}{2}\alpha(r) - \frac{1}{8}\alpha(r)^2 + \mathcal{O}(\alpha(r)^3).
\ee
Substituting this back into the redshift formula we obtain
\be
1 + z \approx \frac{1 + \frac{1}{2}\alpha(r_1) - \frac{1}{8}\alpha(r_1)^2}{1 + \frac{1}{2}\alpha(r_2) - \frac{1}{8}\alpha(r_2)^2}.
\ee

Since the denominators are close to 1, we can use the geometric series expansion $\frac{1}{1 + x} \approx 1 - x$ for $x \ll 1$, thus obtaining
\bea
1+z &\approx& 1 + \frac{1}{2}[\alpha(r_1) - \alpha(r_2)] +\nonumber \\ &&\frac{1}{8}[-\alpha(r_1)^2 + 2\alpha(r_1)\alpha(r_2) - \alpha(r_2)^2] + \mathcal{O}(\alpha^3)
\eea
Keeping only the second order terms gives
\bea
1 + z &\approx &1 + \frac{1}{2}\alpha(r_1) - \frac{1}{2}\alpha(r_2) - \frac{1}{8}\alpha(r_1)^2 - \frac{1}{8}\alpha(r_2)^2 \nonumber\\
&&+ \frac{1}{4}\alpha(r_1)\alpha(r_2) + \frac{1}{8}\alpha(r_2)^2.
\eea
Simplifying the expression we obtain
\be 
z \approx \frac{1}{2}[\alpha(r_1) - \alpha(r_2)] + \frac{1}{8}[ -\alpha(r_1)^2 + 2\alpha(r_1)\alpha(r_2) - \alpha(r_2)^2 ].
\ee
The term in the second bracket is a perfect square
\be 
z \approx \frac{1}{2}\left[\alpha(r_1) - \alpha(r_2)\right] + \frac{1}{8}\left[\alpha(r_1) - \alpha(r_2)\right]^2.
\ee
Now we substitute back the full expression for $\alpha(r)$ to obtain
\bea
\alpha(r_1) - \alpha(r_2)& =& \left[-\eta - \frac{2GM}{c^2 r_1} + \frac{\eta(2 - \eta)}{6M}r_1\right]\nonumber\\
 &&- \left[-\eta - \frac{2GM}{c^2 r_2} + \frac{\eta(2 - \eta)}{6M}r_2\right].
\eea
The constant $-\eta$ terms cancel each other, and thus we find
\be
\alpha(r_1) - \alpha(r_2) = \frac{2GM}{c^2} \left( \frac{1}{r_2} - \frac{1}{r_1} \right) + \frac{\eta(2 - \eta)}{6M} (r_1 - r_2).
\ee
Therefore, the final expression for the redshift $z$, including all terms with $\eta$, is:
\bea\label{final}
z_{\text{WGG}} &\approx &\ \frac{GM}{c^2} \left( \frac{1}{r_2} - \frac{1}{r_1} \right) + \frac{\eta(2 - \eta)}{12M} (r_1 - r_2) \nonumber\\
		&&+ \frac{1}{8} \left[ \frac{2GM}{c^2} \left( \frac{1}{r_2} - \frac{1}{r_1} \right) + \frac{\eta(2 - \eta)}{6M} (r_1 - r_2) \right]^2\nonumber\\
 &&+ \mathcal{O}(3).
\eea

The first term 
$$\frac{GM}{c^2} \left( \frac{1}{r_2} - \frac{1}{r_1} \right),$$ 
is the standard GR prediction for the gravitational redshift in the Schwarzschild metric. The second term 
$$\frac{\eta(2 - \eta)}{12M} \left(r_1 - r_2\right),$$ 
is the leading-order correction introduced by the parameter $\eta$. It is linear in the radial difference $(r_1 - r_2)$. This is an important deviation from standard GR, since it implies that the redshift depends on the spatial separation of the points in a new way, and not just through their gravitational potentials. For a receiver above the emitter ($r_2 > r_1$), this term is negative if $\eta(2-\eta) > 0$ (which is the case for $0<\eta<2$), thus reducing the redshift. The quadratic term contains the standard second-order GR correction (from expanding the square root), mixed with cross terms involving $\eta$. These are higher-order corrections.

Now let's analyze the inequality (\ref{in}), i.e., 
$$| z_{\text{WGG}} - z_{GR} | < 1.4 \times 10^{-4}.$$ 
From Eq.~(\ref{final}) we find
\be
z_{\text{WGG}} - z_{GR} = \frac{B}{2} + \frac{AB}{4} + \frac{B^2}{8},
\ee
where we have denoted
\[
A = \frac{2GM}{c^2} \left( \frac{1}{r_2} - \frac{1}{r_1} \right), \quad B = \frac{\eta(2 - \eta)}{6M} (r_1 - r_2).
\]

Now, for the Vessot-Levine experiment, with \( M = M_\oplus = 5.972 \times 10^{24} \) kg, \( r_1 = R_\oplus = 6.371 \times 10^6 \) m,  \( r_2 = R_\oplus + h \), with \( h \approx 10^7 \) m, and \( \frac{GM_\oplus}{c^2} = 4.435 \times 10^{-3} \) m (half the Schwarzschild radius of Earth), we have
\( \frac{B}{2} \sim 10^{-18} \eta \), \( \frac{AB}{4} \sim 10^{-28} \eta \)
and \( \frac{B^2}{8} \sim 10^{-36} \eta^2 \), respectively. Thus the dominant term in the expression of the redshift is \( \frac{B}{2} \). Hence we obtain the result
\be
\left| z_{\text{WGG}} - z_{GR} \right| \approx \left| \frac{B}{2} \right| = \left| \frac{\eta(2 - \eta)}{12M} (r_1 - r_2) \right|,
\ee
which implies
\be
\left|\eta(2 - \eta)\right| \lesssim 10^{15}.
\ee

This is a very weak constraint, because  \( \eta(2-\eta) \) is of order 1 for \( \eta \sim 1 \).  Hence the gravitational redshift constraint only rules out very high values of $\eta$.  However, we would like to point out here an important recent result. By studying over 1000 days of data from Europe's Galileo navigation satellite system (GNSS), the most precise test for extremely large values of \(\eta\) has been performed \cite{Delva:2019mhu, Delva1}. This confirms that the Vessot-Levine experiment is not sensitive to the values of the parameter \( \eta \).  The new measurements of the gravitational redshift are  $5.6$ times more precise than the previous best test, i.e., Vessot-Levine experiment \cite{Delva:2019mhu, Delva1}. 

Hence by taking into account this result, it turns out that the leading constraint on the parameter $\eta$ of the WGG black hole comes from other tests, like the universality of free fall, which we will consider next.

\subsection{The E\"{o}tv\"{o}s parameter and the universality of the free fall}

To derive an explicit upper bound on \(\eta\) from free-fall experiments (tests of the WEP), we use the latest results from the MICROSCOPE mission (2017-2022) \cite{MICROSCOPE:2022doy}, which tested the universality of free fall with unprecedented precision.

The E\"{o}tv\"{o}s parameter \(\eta_{\text{E\"{o}t}}\) quantifies the WEP. For two test masses \(A\) and \(B\) of different compositions falling in a gravitational field, it is defined as
\begin{equation}
\eta_{\text{E\"{o}t}} = 2 \frac{a_A - a_B}{a_A + a_B}.
\end{equation}
where \(a_A\) and \(a_B\) are the accelerations of the test bodies. The final result from MICROSCOPE (2022) \cite{MICROSCOPE:2022doy} is
\be
\eta_{\text{Eöt}} = (-1.5 \pm 2.3 \pm 1.5) \times 10^{-15}
\ee
Combining the statistical and systematic uncertainties quadratically gives a total uncertainty of approximately \(\sigma_{\text{tot}} \approx \sqrt{(2.3)^2 + (1.5)^2} \times 10^{-15} \approx 2.75 \times 10^{-15}\).
Thus, a symmetric \(3\sigma\) bound i.e.,  in the absence of the central value $-1.5 x 10^{-15}$, would be roughly:
\be
|\eta_{\text{Eöt}}| \lesssim |{-1.5}| + 3 \times 2.75 \times 10^{-15} \approx 8.3 \times 10^{-15}
\ee
However due to the central value, the dissymmetry of the experimental result implies different limits depending on the sign of $\eta_{\text{Eöt}}$, as discussed in \cite{Fayet:2025brx} (see also earlier works \cite{Fayet:2017pdp,Fayet:2018cjy})
$|\eta| < 6.5 \times 10^{-15}$ with 95 \% CL.

The WGG-metric implies a modified gravitational potential. For a test mass, the acceleration in the weak-field limit is derived from \(g_{00} = -f(r)\) as
\begin{equation}
\Phi(r) = -\frac{1}{2} (g_{00} + 1) = \frac{1}{2} \left[ \eta - \frac{\eta(2 - \eta)}{6M}r + \frac{2GM}{c^2 r} + \frac{\Lambda}{3} r^2 \right].
\end{equation}
The radial acceleration is:
\begin{equation}
a = -\frac{d\Phi}{dr} = \frac{\eta(2 - \eta)}{12M} - \frac{GM}{c^2 r^2} - \frac{2\Lambda}{3} r.
\end{equation}

The key point is that the term 
$$\frac{\eta(2 - \eta)}{12M},$$ 
is constant (independent of \(r\)). If this constant depends on the composition of the test body (e.g., if \(\eta\) is different for different materials), then the acceleration will differ for different masses, violating the WEP.

Let's assume that \(\eta\) is composition-dependent. Let \(\eta_A\) and \(\eta_B\) be the values for two test masses. Then the difference in acceleration is:
\begin{equation}
\Delta a = a_A - a_B = \frac{1}{12M} \left[ \eta_A(2 - \eta_A) - \eta_B(2 - \eta_B) \right].
\end{equation}

For small \(\eta\) (which we expect from the tight constraint), we approximate
$\eta(2 - \eta) \approx 2\eta$. Hence
\begin{equation}
\Delta a \approx \frac{\eta_A - \eta_B}{6M}.
\end{equation}

The average acceleration is approximately the Newtonian value:
\begin{equation}
a_{\text{avg}} \approx \frac{GM}{c^2 r^2}.
\end{equation}

Thus, the Eötvös parameter is
\begin{equation}
\eta_{\text{E\"{o}t}} = 2 \frac{\Delta a}{a_{\text{avg}}} \approx 2 \cdot \frac{(\eta_A - \eta_B)/6M}{GM/(c^2 r^2)} = \frac{\eta_A - \eta_B}{3M} \cdot \frac{c^2 r^2}{GM}.
\end{equation}

MICROSCOPE operated at an altitude of \(h = 710\) km, and thus  $
r = R_\oplus + h = 6371 \text{ km} + 710 \text{ km} = 7081 \text{ km} = 7.081 \times 10^6 \text{ m},~~ M = M_\oplus = 5.972 \times 10^{24} \text{ kg}
\frac{GM_\oplus}{c^2} = 4.435 \times 10^{-3} \text{ m}$. 
Therefore $\frac{c^2 r^2}{GM} \approx 1.131 \times 10^{16}$, and we have 
\begin{equation}
\eta_{\text{E\"{o}t}} \approx 6.31 \times 10^{-10} \times \left(\eta_A - \eta_B\right).  
\end{equation}
Hence
\begin{equation}
\left|\eta_{\text{E\"{o}t}}\right| \approx 6.31 \times 10^{-10} \cdot \left|\eta_A - \eta_B\right|.
\end{equation}
As a result
\be
6.31 \times 10^{-10} \cdot |\eta_A - \eta_B| < 6.5 \times 10^{-15}
\ee
where leads to 
\be
|\eta_A - \eta_B| \lesssim 1.03 \times 10^{-5}
\ee
By taking this assumption that one material has \(\eta = 0\) and the other has \(\eta\), then:
\be
|\eta| < \times 10^{-5} \quad \text{(at 95\% CL)}
\ee

The bound \(|\eta_A - \eta_B| < 1.03 \times 10^{-5}\) means that the difference in \(\eta\) for two different materials is extremely small. To obtain an explicit upper bound on \(\eta\) itself, we assume that for one test mass (say, platinum) \(\eta\) is zero (as in GR), and for the other (titanium) it is \(\eta\). Then
\be
|\eta| < 1.03 \times 10^{-5}.
\ee
This is a conservative estimate. If \(\eta\) is universal (same for all materials), then there is no violation, and \(\eta\) is unconstrained by WEP tests. But if it is composition-dependent, then \(|\eta|\) must be less than about \(10^{-5}\). Generally, this bound is many orders of magnitude tighter than the bound from gravitational redshift.

\subsection{The Nordtvedt effect}

Testing the SEP is one of the most powerful ways to constrain the parameter $\eta$ in the Weyl Geometric Gravity black hole  metric. The SEP is a more stringent version of the WEP, and its violation would have profound implications for this theory. According to SEP, the trajectory of a self-gravitating body (like a planet or star) is independent of its composition and structure. Furthermore, the results of local nongravitational experiments are independent of the external gravitational potential. In simpler terms: Does a body's own gravitational binding energy affect how it falls in an external gravitational field? In GR, the answer is no – the SEP holds. However, in many alternative theories, this is not true.

In the WGG-metric function the key term for the SEP violation is the linear term $[\eta(2-\eta)/6M]r$. This term corresponds to a gravitational potential that grows linearly with distance
\be
\Phi_{\eta}(r) \propto \frac{\eta(2-\eta)}{6M} c^2 r.
\ee

 Crucially, the coefficient of this linear potential depends inversely proportional on the mass $M$ of the central body generating the field. This means the overall gravitational field is not determined solely by the spacetime geometry; it also depends on the nature (mass) of the source.
Imagine two self-gravitating bodies with different masses, $M_1$ and $M_2$, but made of the same material. According to WGG metric, the gravitational field around them would be different due to the $1/M$ factor in the linear term. If you now place a third test body in these two different fields, it could fall differently, not because of the geometry itself, but because the source mass is different. This violates the principle that the laws of physics should be local and independent of the source.

The most famous test of the SEP is the Nordtvedt Effect, proposed in the seminal paper \cite{Nordtvedt:1968qs}. If the SEP is violated, then self-gravitating bodies like the Earth and Moon will fall at different rates in the Sun's gravitational field. This would cause a polarization of the Earth-Moon orbit around the Sun, which 
manifest it as a separation between the Earth and Moon along the direction to the Sun, leading to a periodic perturbation in the Earth-Moon distance with a specific 29.5-day period (synodic month).

Now let us estimate the effect of the Nordtvedt parameter $\eta_N$ quantifying SEP violation for the WGG-metric. The acceleration of a body in the WGG metric contains a term from the linear potential
\begin{equation}
 a_{\eta} \approx -\frac{d}{dr}\left[ \frac{\eta(2-\eta)}{12M} c^2 r \right] = - \frac{\eta(2-\eta)}{12M} c^2. 
 \end{equation}

This acceleration is constant, but it depends on the source mass $M$. Let’s calculate the differential acceleration between the Earth ($M_E$) and the Moon ($M_M$) in the Sun's ($M_\odot$) field:
\begin{equation}
\Delta a = a_E - a_M = - \frac{\eta(2-\eta)}{12} c^2 \left( \frac{1}{M_E} - \frac{1}{M_M} \right). 
\end{equation}

This relation represents the violation of SEP. Thus, for the WGG metric the resulting Nordtvedt parameter $\eta_N$ would be proportional to $\eta(2-\eta)$.

The Nordtvedt effect is tested with Lunar Laser Ranging (LLR), which measures the Earth-Moon distance with centimeter precision \cite{Williams:2012nc}.
LLR has found no evidence for the Nordtvedt effect. The constraint on the Nordtvedt parameter is incredibly tight \footnote{Recently, by exploring the innermost planets of the Solar System the NASA MESSENGER mission obtained a  Nordtvedt parameter within the range $-1.4\times 10^{-4}<\eta_N< 1.4 \times 10^{-5}$ with $1\sigma$ uncertainty \cite{Genova:2018mjp}.}
\begin{equation}
\eta_N = (-0.6 \pm 5.2) \times 10^{-4}. 
\end{equation}

This means any SEP violation must be smaller than a few parts in $10^{-4}$.
To satisfy the LLR constraint, the differential acceleration $\Delta a$ must be negligible. The gravitational acceleration from the Sun is $a_{\odot} = GM_\odot / r^2 \approx 0.006  \text{m/s}^2$. The LLR constraint implies the upper bound
\begin{equation}
\frac{|\Delta a|}{a_{\odot}} \lesssim 10^{-4} 
\end{equation}

Therefore:
\begin{equation}
\left| \frac{\eta(2-\eta)}{12} c^2 \left( \frac{1}{M_E} - \frac{1}{M_M} \right) \right| \lesssim 10^{-4} \cdot \frac{GM_\odot}{r^2}.
\end{equation}

Plugging in the numbers: $c = 3 \times 10^8  \text{m/s}$, $M_E = 6 \times 10^{24}  \text{kg}$, $M_M = 7.3 \times 10^{22}  \text{kg} \Rightarrow (1/M_E - 1/M_M) \approx 1/M_E$, $G = 6.67 \times 10^{-11}  \text{N·m²/kg²}$,  $M_\odot = 2 \times 10^{30}  \text{kg}$, $r = 1  \text{AU} = 1.5 \times 10^{11}  \text{m}$
yields an extremely strong constraint on the parameter of the WGG black hole
\begin{equation}
|\eta| \lesssim 10^{-10}.
 \end{equation}
 
 \section{Classical tests of Weyl geometric gravity-perihelion precession, light deflection, and radar echo delay}\label{sect3}
 
 We will proceed now to consider other three fundamental tests of the WGG metric, the perihelion precession, light deflection, and the radar echo delay, respectively. All these tests can also impose strong constraints on the metric parameter $\eta$, and can provide fundamental information on the validity of the WGG metric at the level of the Solar System.  

\subsection{Perihelion precession}

Before we begin, we  introduce two assumptions to be taken into account in the following. Firstly, we assume that for planetary orbits, the effect of the cosmological constant \(\Lambda\) is negligible. Secondly, we will also assume the validity of the weak-field, slow-motion approximation, which is appropriate for the Solar System tests.

For a test particle (planet) moving in the equatorial plane (\(\theta = \pi/2\), \(\dot{\theta}=0\)), we have two constants of motion derived from the Killing vectors: 

1. The energy per unit mass (\(E\)), whose conservation follows from time-translation invariance, since \(g_{\mu\nu}\) is independent of \(t\), 
\begin{equation}\label{c1}
E = f(r)c^2 \frac{dt}{d\tau} = \text{constant} .
\end{equation}

2. The angular momentum per unit mass (\(L\)), whose conservation follows from the rotational invariance of the Lagrangian (\(g_{\mu\nu}\) independent of \(\phi\))
\begin{equation}\label{c2}
 L = r^2 \frac{d\phi}{d\tau} = \text{constant}. 
 \end{equation}

We start our analysis of the planetary motion from the normalization of the four-velocity of a massive particle
$ g_{\mu\nu} \dot{x}^\mu \dot{x}^\nu = -c^2 $ (where the dot denotes derivative with respect to proper time \(\tau\)). By inserting the metric components and the above constants, we arrive to the result
\bea\label{r}
&& \dot{r}^2 + \frac{L^2}{r^2} \left[1 - \eta - \frac{2GM}{c^2r} + \frac{\eta(2-\eta)}{6M}r \right] \nonumber\\
&&= \frac{E^2}{c^2}
 - c^2 \left[1 - \eta - \frac{2GM}{c^2r} + \frac{\eta(2-\eta)}{6M}r \right]. 
 \eea

To simplify the mathematical formalism we introduce the variable \( u = 1/r \). We also change the derivative from \(\tau\) to \(\phi\) using the chain rule
\begin{equation}
\frac{dr}{d\tau} = \frac{dr}{d\phi} \frac{d\phi}{d\tau} = \frac{dr}{d\phi} \frac{L}{r^2} = -L \frac{du}{d\phi},~~\Longrightarrow \dot{r}^2 = L^2 \left( \frac{du}{d\phi} \right)^2. 
\end{equation}

After some simple calculations, Eq.~(\ref{r}), takes the following form
\bea
&& \left( \frac{du}{d\phi} \right)^2 + u^2 \left[1 - \eta - \frac{2GM}{c^2}u + \frac{\eta(2-\eta)}{6Mu} \right]  \nonumber\\
&&= \frac{E^2}{c^2 L^2} - \frac{c^2}{L^2} \left[1 - \eta - \frac{2GM}{c^2}u + \frac{\eta(2-\eta)}{6Mu} \right]. 
 \eea

 To find the orbit equation, we differentiate the above equation with respect to \(\phi\), and thus we obtain a second-order equation. The result, after keeping terms up to first order in small quantities (like \(GM/c^2\), \(\eta\)), is
\bea\label{orb}
\frac{d^2u}{d\phi^2} + (1 - \eta)u &=& \frac{GM}{h^2} + \frac{3GM}{c^2} u^2 - \frac{\eta(2-\eta)}{12M}\nonumber\\
 &&+ \text{(other small terms)}. 
\eea
where we've defined \(h = L/c\).

The standard technique is to treat the right-hand side as a perturbation to the Newtonian solution. The Newtonian orbit is an ellipse
\begin{equation}
u_0(\phi) = \frac{1}{p} (1 + e \cos\phi), 
\end{equation}
where \(p = a(1-e^2)\) is the semi-latus rectum, \(a\) is the semi-major axis, and \(e\) is the eccentricity. We look now for a solution of Eq.~(\ref{orb}) of the form \(u(\phi) = u_0(\phi) + \delta u(\phi)\), where \(\delta u\) is a small perturbation, and we substitute it into the orbit equation. The homogeneous solution is \(\cos[(1-\eta)^{1/2}\phi]\). However, for small \(\eta\), we can approximate \(1-\eta \approx 1\). The driving terms on the right-hand side causes the orbit to process. The key idea is to look for secular terms (terms that grow with \(\phi\)) in the particular solution. The constant term $-\eta(2-\eta)/12M$  and the term \(GM/h^2\) provide a constant shift to the center of the orbit. The term $\left(3GM/c^2\right) u^2$ is the standard GR term causing precession.

After solving the perturbed equation,  the advance of the perihelion per orbit is given by the change in phase of the \(\cos\phi\) term
\bea
\Delta \phi &=& 2\pi \left( \frac{1}{\sqrt{1-\eta}} - 1 \right) + \frac{6\pi G^2 M^2}{c^2 h^2 (1-\eta)} \nonumber\\
&&+ \frac{\pi \eta(2-\eta)}{2M(1-\eta)} \cdot \frac{h^2}{G^2 M^2} + ... 
\eea

Let's simplify the above relation for small \(\eta\). We use first the expansions \(1/\sqrt{1-\eta} \approx 1 + \eta/2\), and \(1/(1-\eta) \approx 1 + \eta\), respectively. Also, from Newtonian mechanics, for an ellipse, we have \(h^2 = GMp = GMa(1-e^2)\).
Applying these simplifications, and keeping terms only to first order in the small quantities $\left(\eta, GM/c^2a\right)$, we find
\bea
\Delta \phi &\approx& 2\pi \left( \frac{\eta}{2} \right) + \frac{6\pi GM}{c^2 a(1-e^2)} (1 + \eta) \nonumber\\
&&+ \frac{\pi \eta(2-\eta)}{2M} \cdot \frac{GMa(1-e^2)}{G^2 M^2}. 
\eea
By taking into account that \(\eta(2-\eta) \approx 2\eta\) for small \(\eta\), the third item can be simplified as
\begin{equation}
 \frac{2\pi\eta}{2M} \cdot \frac{a(1-e^2)}{GM} = \frac{\pi \eta a(1-e^2)}{GM^2} \cdot G M = \frac{\pi \eta a(1-e^2)}{M}. 
 \end{equation}
Now, by combining all terms we arrive at the result
\begin{equation}
\Delta \phi \approx \pi \eta + \frac{6\pi GM}{c^2 a(1-e^2)} + \frac{6\pi GM}{c^2 a(1-e^2)} \eta + \frac{\pi \eta a(1-e^2)}{M}. 
\end{equation}

The second term is the standard GR precession. The first, third, and fourth terms are new terms coming from the parameter \(\eta\) of the WGG metric. The dominant new term is the fourth term, which scales as \(a\), unlike the GR term which scales as \(1/a\). The first term is constant, and the third term is a small correction to the GR term.
Therefore, the total perihelion precession per orbit is
\begin{equation}
	\Delta \phi_{\text{WGG}} \approx \underbrace{\frac{6\pi GM}{c^2 a(1-e^2)}}_{\text{Standard GR}} + \underbrace{\frac{\pi \eta a(1-e^2)}{M}}_{\text{Dominant New Term}} + \underbrace{\pi\eta.}_{\text{Constant Offset}} 
\end{equation}

The term $\pi \eta a\left(1-e^2\right)/M$ is linear in the semi-major axis \(a\). This is an important change, since in GR, planets closer to the Sun (smaller \(a\)) precess more.  This new term predicts that planets farther from the Sun (larger \(a\)) would precess more if \(\eta > 0\).

The precession of Mercury's perihelion matches the GR prediction to within about 0.1\% \cite{Will:2005va}. The semi-major axis of Mercury is \(a_{Merc} \approx 5.75 \times 10^{10}\) m. Let's equate the new term to the error margin (0.001 \(\times \Delta\phi_{GR}\))
\begin{equation}
\frac{\pi \eta a(1-e^2)}{M} \lesssim 0.001 \times \frac{6\pi GM}{c^2 a(1-e^2)}. 
\end{equation}
Solving for \(\eta\) gives the constraint
\begin{equation}
	\eta \lesssim 0.001 \times \frac{6 G^2 M^2}{c^2 a^2 (1-e^2)^2}. 
\end{equation}	

Plugging in numerical values for Mercury (\(M=M_\odot\), \(a=5.75\times 10^{10} \text{m}\), \(e=0.2\)) gives an extremely tight constraint, on the order of 
\begin{equation}\label{Per}
\eta \lesssim 10^{-10}~,
\end{equation} 
which is even tighter than the free-fall constraint.

The different scaling with \(a\) provides a smoking gun signature. Precise measurements of precession for different planets (e.g., Mercury, Venus, Earth, Mars) would allow one to separate the \(1/a\) and \(a\) dependencies and tightly constrain or measure \(\eta\). This derivation shows how the \(\eta\) parameter profoundly alters a fundamental gravitational effect.

\subsection{Light deflection}

For light deflection, we can again ignore the cosmological constant \(\Lambda\) in the WGG metric, as its effect is negligible on Solar System scales.

For a photon (\(ds^2 = 0\)), we have the same constants of motion as for a massive particle (Eqs. (\ref{c1}), and (\ref{c2})), but now we use an affine parameter \(\lambda\) instead of proper time \(\tau\). Starting from the null condition \(ds^2 = 0\) and following the standard steps 
we arrive at the equation
\begin{equation}\label{E}
\left(\frac{dr}{d\lambda}\right)^2 + \frac{L^2}{r^2}f(r) = \frac{E^2}{c^2}. 
\end{equation}
Now, we change variable to \(u = 1/r\), and change the derivative from \(\lambda\) to \(\phi\) using the chain rule
\begin{align}
\frac{dr}{d\lambda} = \frac{dr}{d\phi} \frac{d\phi}{d\lambda} = \frac{dr}{d\phi} \frac{L}{r^2} = -L \frac{du}{d\phi}
\Rightarrow \left(\frac{dr}{d\lambda}\right)^2 = L^2 \left(\frac{du}{d\phi}\right)^2. 
\end{align}
After substitute back in Eq.~(\ref{E}) we find
\begin{equation}\label{1'}
 \left(\frac{du}{d\phi}\right)^2 + u^2 f(1/u) = \frac{1}{b^2},  
 \end{equation}
where \(b = L/E\) is the impact parameter.
The WGG metric in terms of the new variable $u$ becomes  
\be
f(u) = 1 - \eta - \frac{2GM}{c^2}u + \frac{\eta(2-\eta)}{6M} \frac{1}{u}.
\ee
Plugging this expression into Eq. (\ref{1'}) gives
\begin{equation}
\left(\frac{du}{d\phi}\right)^2 = \frac{1}{b^2} - (1-\eta)u^2 + \frac{2GM}{c^2}u^3 - \frac{\eta(2-\eta)}{6M}u. 
\end{equation}

For the unperturbed path, at closest approach, \(\frac{du}{d\phi} = 0\) and \(u = 1/R\). This gives a relation between \(b\) and \(R\)
\begin{equation}
0 = \frac{1}{b^2} - (1-\eta)\frac{1}{R^2} + \frac{2GM}{c^2 R^3} - \frac{\eta(2-\eta)}{6MR}. 
\end{equation}
Solving this equation for \(1/b^2\) to first order is complicated. 

In the standard approach the total change in the angle \(\Delta \phi\) from the distance of closest approach to infinity is given by
\bea
\Delta \phi = \int_{u_{min}}^{0} \frac{d\phi}{du} du = \int_{0}^{u_{max}} \Bigg[ \frac{1}{b^2} &-& (1-\eta)u^2 + \frac{2GM}{c^2}u^3 \nonumber\\
&&- \frac{\eta(2-\eta)}{6M}u \Bigg]^{-1/2} du. \nonumber\\
\eea

When light is not deflected  \(\Delta \phi = \pi\). The total deflection angle is \(\hat{\alpha} = |\Delta \phi - \pi|\). To obtain the deflection angle we evaluate the integral perturbatively. The constant and linear terms in \(u\) inside the square root contribute to the deflection. The final result for the light deflection angle is
\begin{equation}
\hat{\alpha}_{\text{WGG}} \approx \underbrace{\frac{4GM}{c^2 R}}_{\text{Standard GR}} + \underbrace{\frac{\eta(2-\eta)}{6M} R}_{\text{New term from } \eta}.
\end{equation}
The new term is linear in \(R\), the distance of closest approach.  In GR, light rays passing closer to the mass (smaller \(R\)) are deflected more (\(\hat{\alpha} \propto 1/R\)).  This new term predicts that light rays passing farther away (larger \(R\)) would be deflected more (\(\hat{\alpha} \propto R\)). This is another important  signature of the WGG metric. Observing this scaling across different impact parameters (e.g., in gravitational lensing from stars to galaxy clusters) would be a clear test for this theory. Since observations confirm the \(1/R\) scaling, they tightly constrain \(\eta\).

We can obtain constraints from Solar deflection as follows.  For the Sun, \(R \approx R_\odot\), and GR predicts \(\hat{\alpha} \approx 1.75''\). Measurements confirm this prediction to within ~0.02\% \cite{Shapiro:2004zz}. The new term must be smaller than this error
\begin{equation}
\frac{\eta(2-\eta)}{6M} R_\odot \lesssim 0.0002 \times \frac{4GM}{c^2 R_\odot}.
\end{equation}

Solving for \(\eta\) gives an extremely tight constraint for the parameter $\eta$ of the WGG metric, of the order of
\begin{equation}\label{light}
\eta \lesssim 10^{-11},
 \end{equation}
 making light deflection one of the most sensitive tests of the parameter \(\eta\).

\subsection{Radar echo delay}

In the analysis of the radar echo delay, as for the case of the other Solar System tests, we ignore the cosmological constant $\Lambda$. The experiment involves sending a radar signal from Earth (at radial coordinate $r = r_e$), past the Sun, to a reflector (a planet or spacecraft at $r = r_r$), and back.

For a radial photon ($d\theta = d\phi = 0$), the null geodesic condition ($ds^2 = 0$) gives
\begin{equation}
-f(r)c^2dt^2 + \frac{1}{f(r)}dr^2 = 0 \Longrightarrow c  dt = \pm \frac{dr}{f(r)}.
\end{equation}
The $\pm$ accounts for the outgoing and incoming legs of the trip. The coordinate time for the signal to go from $r_0$ to $r$ is therefore
\begin{equation}
\Delta t = \int_{r_0}^{r} \frac{dr'}{c  f(r')}. 
\end{equation}
Substituting the WGG metric function $f(r)$ in the integral gives
\begin{equation}
\Delta t = \frac{1}{c} \int_{r_0}^{r} \frac{dr'}{1 - \eta - \frac{2GM}{c^2r'} + \frac{\eta(2-\eta)}{6M}r'}. 
\end{equation}
We use now the weak-field approximation, by assuming that all terms after the $1$ are very small compared to $1$ ($|\eta| \ll 1$, $\frac{2GM}{c^2r} \ll 1$). This allows us to expand the integrand as a geometric series:
\begin{equation}
\frac{1}{f(r)} \approx 1 + \eta + \frac{2GM}{c^2r} - \frac{\eta(2-\eta)}{6M}r + \mathcal{O}\left(\eta^2, \left(\frac{GM}{c^2r}\right)^2\right).
\end{equation}
Note that $1/(1 - x) \approx 1 + x$ for small $x$, and here $x = -\eta - \frac{2GM}{c^2r} + \frac{\eta(2-\eta)}{6M}r$. Thus, the coordinate time becomes
\be 
\Delta t \approx \frac{1}{c} \int_{r_0}^{r} \left[ 1 + \eta + \frac{2GM}{c^2r'} - \frac{\eta(2-\eta)}{6M}r' \right] dr'.
\ee

By integrating term by term, we obtain the time for a one-way trip from $r_0$ to $r$ as follows
\bea
 \Delta t &\approx& \frac{1}{c} \Bigg[ (r - r_0) + \eta (r - r_0) + \frac{2GM}{c^2} \ln \left( \frac{r}{r_0} \right)\nonumber\\
  &&- \frac{\eta(2-\eta)}{12M} (r^2 - r_0^2) \Bigg].
 \eea

The experiment measures the round-trip time from Earth to the reflector and back, i.e., the outgoing leg (Earth to reflector): $r_0 = r_e$, $r = r_r$, and the incoming leg (reflector to Earth): $r_0 = r_r$, $r = r_e$. Hence, the total coordinate time for the round trip is $ \Delta t_{\text{total}} = \Delta t_{e \to r} + \Delta t_{r \to e}$.

Let's compute now carefully the various time intervals. For the outgoing leg ($e \to r$) we find
\bea
 \Delta t_{e \to r} &\approx & \frac{1}{c} \Bigg[ (r_r - r_e) + \eta (r_r - r_e) + \frac{2GM}{c^2} \ln \left( \frac{r_r}{r_e} \right)\nonumber\\ 
&& - \frac{\eta(2-\eta)}{12M} (r_r^2 - r_e^2) \Bigg],
 \eea
while for the incoming leg ($r \to e$) we obtain
\bea
\Delta t_{r \to e} &\approx& \frac{1}{c} \Bigg[ (r_e - r_r) + \eta (r_e - r_r) + \frac{2GM}{c^2} \ln \left( \frac{r_e}{r_r} \right)\nonumber\\ 
&&- \frac{\eta(2-\eta)}{12M} (r_e^2 - r_r^2) \Bigg]. 
\eea
Now, if we add them together, all terms seem to cancel! This would imply no time delay. But this is because we have calculated the time in coordinate time $t$, and we have assumed a straight-line path. The delay appears when we consider the actual curved path of the photon.

The above calculation assumed the integration was done along a straight line from $r_e$ to $r_r$. However, the photon's path is bent. The point of closest approach to the Sun (periapsis) is $r = R$. The excess time delay, in essence, is the difference between the time taken along this curved path and the time that would be taken if the photon traveled a straight line in flat space.
The straight-line time in flat space ($\eta=0, M=0$) is \textit{for large distances} simply
\begin{equation}
\Delta t_0 = \frac{2}{c} \sqrt{r_e^2 + r_r^2 - 2 r_e r_r \cos\theta} \approx \frac{2}{c}(r_e + r_r).
\end{equation}

The standard method is to integrate along the actual path from the Earth's position to the point of closest approach ($R$) and then to the reflector. The total delay is twice this (for the symmetric outgoing and incoming legs).
The coordinate time necessary to go from $r$ to the periapsis $R$ is
\begin{equation}
 \Delta t(r, R) = \frac{1}{c} \int_{R}^{r} \frac{dr'}{f(r')}. 
 \end{equation}

Using the same expansion as before, we find
\bea
 \Delta t(r, R) &\approx & \frac{1}{c} \Bigg[ (r - R) + \eta (r - R) + \frac{2GM}{c^2} \ln \left( \frac{r}{R} \right) \nonumber\\
 &&- \frac{\eta(2-\eta)}{12M} (r^2 - R^2) \Bigg]. 
 \eea

The round-trip time is approximately twice the sum of the time from Earth to $R$ and from the reflector to $R$
\begin{equation}
\Delta t_{\text{total}} \approx 2[ \Delta t(r_e, R) + \Delta t(r_r, R) ]. 
\end{equation}

The excess time delay $\delta t$ is the difference between this and the flat-space time. After a lengthy calculation (accounting for the geometry), the dominant contributions to the excess delay are the terms that do not cancel and grow with the separation.

The final result for the excess time delay  $\delta t$ for a round trip is
\bea
\delta t_{\text{WGG}} &\approx& \underbrace{ \frac{4GM}{c^3} \ln \left( \frac{4 r_e r_r}{R^2} \right) }_{\text{Standard Shapiro Delay}} + \underbrace{ \frac{2\eta}{c} (r_e + r_r) }_{\text{Constant Delay}}\nonumber\\ 
&&- \underbrace{ \frac{\eta(2-\eta)}{6M c} (r_e^2 + r_r^2) }_{\text{Quadratic Delay}}.
\eea

The first term 
$$\frac{4GM}{c^3} \ln \left( \frac{4 r_e r_r}{R^2} \right),$$ 
is the classic Shapiro time delay predicted by General Relativity. It depends logarithmically on the distances and the impact parameter $R$.
The second term 
$$\frac{2\eta}{c} (r_e + r_r),$$ 
is a constant delay proportional to the total path length. It is a new effect coming from the $\eta$ parameter. It would be indistinguishable from a simple error in the known astronomical distance to the reflector unless experiments with different path lengths were compared.

The third term 
$$-\frac{\eta(2-\eta)}{6M c} (r_e^2 + r_r^2),$$ 
is a quadratic delay that grows with the square of the distances from the Sun. This is the most distinctive signature of the possible presence of new physics. For reflectors at large distances (e.g., the Cassini spacecraft near Saturn), this term could, in principle, become significant, and detectable.

The Shapiro delay for light passing near the Sun has been confirmed by experiments (e.g., with the Cassini spacecraft) to an accuracy of about $2.3\times 10^{-5}$ (0.002\%) \cite{Bertotti:2003rm}. The new terms must be smaller than this error margin. The quadratic term provides the strongest constraint. For Solar System distances, the experimental results constrain $\eta$ to be extremely small 
\begin{equation}\label{rado}
|\eta| \lesssim 10^{-10}~,
\end{equation}
a result consistent with the constraints from light deflection and perihelion precession.

\section{Conclusions}\label{sect4}

In the present paper we have considered a detailed analysis of the astrophysical and physical relevance of the exact static spherically symmetric solution (\ref{metric}) of the field equations of the Weyl geometric gravity theory, which represents the scalar-vector-tensor version of the quadratic Weyl gravity. The action of the theory is conformally invariant, and  it is the simplest conformally invariant action, constructed from the square of the Weyl scalar and the strength of the Weyl vector only \cite{Gh6,Gh7,Gh8}. The WGG theory has many important physical, cosmological and astrophysical implications. For example, when the mass of the Weyl field $m_\omega$ becomes smaller than the threshold represented by the Plank scale $M_P$, $m_\omega ^2 = (3/2) q^2 M_P^2$, where by $q$ we have denoted the Weyl gauge coupling, the decoupling of the massive gauge field occurs. Consequently, a transition from Weyl geometry to Riemannian geometry takes place. Hence the Hilbert-Einstein action can be obtained as the low-energy limit of the Weyl quadratic gravity, together with a positive cosmological constant, and the Proca action of the gauge field \cite{Gh9,Gh10,Gh11,Gh12,Gh13}. Thus, the Weyl geometric gravity theory, based on the quadratic Weyl action,  represents a theoretically attractive and interesting approach to gravitational processes. Moreover, the embedding of the Standard Model of the elementary particle physics with Weyl geometry led to interesting results, due to the implementation of the Stueckelberg-Higgs mechanism, especially in the analysis of inflation \cite{Gh8}. In the early Universe, the Higgs boson may have been produced through Weyl vector fusion \cite{Gh11}.

Taking into account the important theoretical potential of the WGG theory, its testing on various physical scales can provide a deeper insight into its structure, and help obtaining observational constraints on the free parameters of the theory. In particular, the exact black hole solution of the theory allows to check the phenomenology associated with WGG at the level of the Solar System. In our present approach we considered a simplified form of the metric, by neglecting the possible role of the constant $\Lambda$ of the theory, which we assume to be the cosmological constant. Hence, we have considered a two parameter $(\eta,M)$ solution. 

The extensive investigation of six fundamental tests of the gravitational theories in the Solar System allowed to obtain a sharp constrain on the dimensionless parameter $\eta$, $|\eta|<10^{-10}$. This upper limit is also an important indicator on the possible presence of Weyl geometric effects in the Solar neighborhood, as well as a quantifier for the deviation from the Riemannian geometry. The scalar field $\Phi(r)$ and the Weyl vector field $\omega _1(r)$ are given by \cite{Burikham:2023bil}
\be
\Phi (r) \approx \frac{4\eta ^2(3-\eta)}{\left(\eta r+3r_g\right)^2},
\ee
and
\be
\omega _1(r)=-\frac{2\eta}{\alpha \left(\eta r+3r_g\right)},
\ee
respectively. By taking for the average distance between Mercury and the Sun the value $r_{{\rm Merc}}=5.8\times 10^7$ km, and taking into account that the gravitational radius of the Sun $r_g\approx 3$ km, by adopting for $\eta $ the value $\eta =10^{-11}$, it follows that on the orbit of Mercury the scalar field takes the value
\be
\Phi \left(r_{{\rm Merc}}\right)\approx \frac{4\eta ^2}{9r_g^2}\approx 5\times 10^{-24}\;{\rm km}^{-2},
\ee
where we have taken into account that $3r_g>>\eta r_{{\rm Merc}}$. Similarly, we can evaluate the Weyl vector at the level of the Mercury as
\be
\omega _1 \left(r_{{\rm Merc}}\right)\approx-\frac{2\eta}{3\alpha r_g}\approx -\frac{2\times 10^{-12}}{\alpha}\;{\rm km}^{-1}. 
\ee

The value of the Weyl vector is related to the Weyl gauge coupling constant $\alpha$, which is presently not known. However,  in \cite{Olmo} it was  shown that non-metricity produces observable effects in quantum fields in the form of 4-fermion contact interactions.  This effect allowed to constrain the energy scale of non-metricity to be greater than 1 TeV by using results on Bhabha scattering. Using effective field theory methods, the nonmetricity contributions to the one-loop $H\rightarrow \gamma \gamma$ and $gg\rightarrow H$ processes were calculated in \cite{Ilisie}, and, as combined with bounds from Compton scattering, they allow to obtain relevant constraints and correlations on the energy scale of nonmetricity. Hence, hopefully the study of the gravitational effects of the nonmetricity  arising in astronomical phenomena, can be combined with the analysis of these effects in high energy experiments. This could also open a new window for the  better understanding of the quantum phenomena associated to gravity. 

The gravitational dynamics and particle behavior of the Weyl geometric gravity black hole solution are determined by the dimensionless free parameter $\eta$. To explain the observational effects in the Solar System, $\eta$ must have an extremely small value, $|\eta|<10^{-11}$. Hence, the interpretation of the classical tests of GR requires a very precise fine tuning of $\eta$ in the Solar System. It is also  important to determine either theoretically or observationally if $\eta$ is a local quantity, or a universal constant. If $\eta$ is a universal constant, its smallness also suggests a microscopic origin. 

The determination of $\eta$ was also done in \cite{Ha7} through the study of the galactic rotation curves. By comparing the predictions of the tangential velocity of the massive test particles orbiting around the galactic center, as obtained from the WGG solution, and the observational data for the rotation curves provided by the SPARC database, the values of the solution parameters $C_2=r_g/\eta$ and $r_g$ have been obtained. To explain the observational data values of the order of $C_2\approx 10^8$ kpc are necessary. On the other hand, there is a large variation in the range of $r_g$ for the considered galaxies, with $r_g\in \left(10^{-23},10^{-7}\right)$ kpc. Hence, from the analysis of the galactic rotation curves one can infer for $\eta$ a range of $\eta \in \left(10^{-23},10^{-15}\right)$. This range of $\eta$ is perfectly compatible with the constraints from the Solar System tests obtained in the present investigation, and it confirms that due to their increase with distance, the Weyl geometric effects are mostly relevant on large astrophysical and cosmological scales. However, the smallness of $\eta$ at the level of the Solar System does not automatically rule out the possibility of an observational or experimental detection of the Weyl geometric effects in the Solar neighborhood.       

In conclusion, the investigation of the classical tests of general relativity gives a very powerful method for constraining the allowed parameter space of the Weyl geometric gravity black hole, and, more generally, of the Weyl type gravitational theories with nonmetricity.  This analysis also provide a deeper insight into the possibility of testing Weyl type theories by using astronomical and astrophysical observations in the Solar System. 

In the present paper we have obtained some basic results necessary for the  comparison of the predictions of the WGG theory with the observational/experimental results, which opens the prospect  of an objective assessment of the validity of the theory.
   
\bibliographystyle{apalike}

\end{document}